\documentstyle[preprint,aps,epsfig]{revtex}

\title{$N-d$ scattering including electromagnetic forces}
\author{A. Kievsky, M. Viviani, and L.E. Marcucci}
\address{Istituto Nazionale di Fisica Nucleare, and 
Dipartimento di Fisica, Universit\`{a} di Pisa, Via Buonarroti 2, I-56100 Pisa,
Italy}

\def\bm{\boldmath}
\def\x{{\bf x}}
\def\y{{\bf y}}
\begin{document}

\maketitle

\begin{abstract}
The electromagnetic potential consisting in the Coulomb plus the magnetic 
moment interactions between two nucleons is studied in nucleon-deuteron
scattering. For states in which the relative $N-d$ angular momentum $L$
has low values the three--nucleon problem has been solved using the
correlated hyperspherical harmonic expansion basis. For states
in which the angular momentum $L$ has large values,
explicit formulae for the nucleon-deuteron 
magnetic moment interaction are derived and used to calculate
the corresponding $T$-matrices in Born approximation.
Then, the transition matrices describing $N-d$
elastic scattering have been derived
including an infinite number of partial waves as required by the
$1/r^3$ behavior of the magnetic moment interaction. Appreciable effects are
observed in the vector analyzing powers at low energies.
The evolution of these effects by increasing the collision energy
is examined.
\end{abstract}

\section{Introduction}

The study of the magnetic moment interaction (MM) in the two-nucleon ($2N$)
system has been subject of many investigations (see 
Refs.~\cite{knutson78,stoks90} and references there in). Although the
intensity of this interaction is very small compared to the nuclear interaction,
its long range behavior produces significant effects in nucleon--nucleon
($NN$) scattering.
Almost all modern $NN$ potentials have been constructed considering
the electromagnetic (EM) interaction used in the Nijmegen partial-wave analysis
which includes the MM interaction between the two spin-$\frac{1}{2}$
particles as well as corrections to the $p-p$ Coulomb potential as two-photon
exchange, Darwin-Foldy and vacuum polarization terms.
When $2N$ scattering observables are computed with one of
these potentials the long range behavior of the EM interaction
implies an infinite sum in the partial-wave series.
For the particular case of
the MM interaction, in Refs.~\cite{knutson78,stoks90}
it has been shown how to sum analytically these infinite series for $p-p$ and
$n-p$ scattering. Important effects of the MM interaction
has been observed in both $n-p$ and $p-p$ vector analyzing powers 
at low energies.

Due to the fact that $2N$ potentials are constructed by fitting
the $NN$ available data, the three-nucleon ($3N$) system is the simplest one
in which these potentials can be used to make predictions. However,
in the description of the $3N$ continuum the MM interaction and corrections
to the Coulomb potential has been systematically disregarded. This omission has 
been justified in the past by the intrinsic difficulties in solving the nuclear 
problem. At present, the $3N$ continuum is routinely solved by different
techniques making possible the treatment of those electromagnetic terms
beyond the Coulomb interaction.

In the present paper we study $N-d$ elastic scattering including Coulomb plus
MM interactions. Previous description of this process without
considering the MM interaction has been performed by the authors
using a technique based on the Kohn variational principle 
(KVP)~\cite{kvr2001,viviani2000} and expanding the scattering wave function in
terms of the correlated hyperspherical harmonics basis~\cite{phh1,phh2}.
Following these works we perform a partial-wave decomposition of the scattering
process. For states with low values of the relative orbital angular momentum $L$ of the 
projectile and the target, the process is studied
by solving the complete $3N$ problem with the Hamiltonian of
the system containing nuclear plus Coulomb plus MM interactions. 
For states with $L$ values sufficiently high, 
the centrifugal barrier prevents a close approach of the projectile
to the target. So, the collision can be considered peripheral and treated as a 
two-body process. Furthermore, in these states only the EM interaction gives
appreciable effects and the corresponding
scattering amplitudes can be calculated in Born approximation. The value of $L$
at which the treatment of the problem changes from a three-body description
to a two-body description is to some extent arbitrary and could be different
at different energies. In practice it can be taken equal to the maximum
$L$ value considered when the problem is solved neglecting the MM
interaction.

We apply this procedure to calculate the $3N$ vector analyzing powers where
the main effects of the MM interaction can be observed. For 
$p-d$ scattering a sizable increase in $A_y$ and $iT_{11}$ has been obtained
at low energies which is, however, insufficient to explain the usual 
underestimation produced by modern $NN$ forces~\cite{puzzle,report}. 
Other observables as the differential cross section and the tensor analyzing
powers suffer minor modifications, of the order of $1$\% or less. 
For $n-d$ scattering a pronounced effect at very small angles is observed.
In fact,
the scattering amplitude has a term $\sin{\theta}/(1-\cos{\theta})$ which
diverges for $\theta\rightarrow 0$ similarly to the $n-p$ case~\cite{stoks90}.
The experimental observation of this divergence is problematic since it
occurs at extreme forward angles (a fraction of degree). This is different from
the $p-d$ case in which the Coulomb divergence dominates in that region.
Regarding the vector analyzing powers, the MM interaction tends
to slightly flatten the $n-d$ $A_y$ around the peak and to produce 
a pronounced dip structure at small scattering angles.

The importance of the EM interaction in the description of $N-d$ scattering
decreases as the energy of the process increases. Around $E_{lab}=16$ MeV
the improvement given by the MM interaction at the peak of $A_y$ and $iT_{11}$
for $p-d$ scattering is already less than $5$\%. On the other hand
Coulomb effects are important below $E_{lab}=30$ MeV~\cite{kvr2001}.
Here we show that at $E_{lab}=65$ MeV they are considerably reduced in most
of the observables with the exception of $T_{21}$ where still some effects
can be observed. This analysis will serve to justify the application of standard $n-d$
calculations to the description of $p-d$ scattering at high 
energies~\cite{witala1}.

The paper is organized as follows. In Section II the $N-d$ MM interaction
is given. The corresponding $T$-matrices are calculated in Born approximation
for both $n-d$ and $p-d$ scattering and final forms for the transition
matrices are given. In Section III the transition from a $3N$ description
to a $2N$ description is discussed. It is shown that the $3N$ $T$-matrix
tends to the $2N$ $T$-matrix as the value of $L$ increases. In Section IV
the vector analyzing powers are calculated including the MM interaction
and compared to the available data. The differences between the theory
and the experiments around the peak of the observables
are analyzed. In Section V we present our conclusions. In the Appendix
the $n-d$ as well as the $p-d$ MM interactions as two
distinctive particles are derived.

\section{Magnetic Moment Interaction}

Following the notation used in the determination of the Argonne $v_{18}$
(AV18) potential~\cite{av18}, 
all modern $NN$ potentials can be put in the general form

\begin{equation}
   v(NN)=v^{EM}(NN)+v^\pi(NN)+v^R(NN)\ .
\end{equation}

The short range part $v^R(NN)$ of these interactions
includes a certain number of parameters (around 40), 
which are determined by a fitting procedure to the $NN$ scattering data and
the deuteron binding energy (BE), whereas the long range part 
reduces to the one-pion-exchange potential $v^\pi(NN)$
and the electromagnetic potential $v^{EM}(NN)$. 

The AV18 potential includes the same $v^{EM}(NN)$ used in the Nijmegen 
partial-wave analysis except for short-range terms and finite size
corrections. The $v^{EM}(pp)$
consists of the one- and two-photon Coulomb terms plus the
Darwin-Foldy term, vacuum polarization and MM interactions.
The $v^{EM}(np)$ interaction includes a Coulomb term due to the neutron charge
distribution in addition to the MM interaction. Finally,
$v^{EM}(nn)$ is given by the MM interaction only. All
these terms take into account the finite size of the nucleon charge
distributions. Explicitly the two--nucleon magnetic moment interaction 
in the center of mass reference frame reads:

\begin{eqnarray}
v_{MM}(pp) & = & -{\alpha\over 4M^2_p}\mu^2_p\left[{2\over 3}F_\delta(r)
  \mbox{\bm$\sigma$}_i\cdot\mbox{\bm$\sigma$}_j+{F_t(r)\over r^3}S_{ij}\right]
  \nonumber \\
 && -{\alpha\over 2 M^2_p}(4\mu_p-1){F_{ls}(r)\over r^3}\bf L\cdot \bf S  
 \;\; , \label{mmpp} \\
v_{MM}(np) & = & -{\alpha\over 4 M_n M_p}\mu_n\mu_p\left[{2\over 3}F_\delta(r)
  \mbox{\bm$\sigma$}_i\cdot\mbox{\bm$\sigma$}_j+{F_t(r)\over r^3}S_{ij}\right]
  \nonumber \\
 && -{\alpha\over 2 M_nM_{np}}\mu_n{F_{ls}(r)\over r^3}
  (\bf L\cdot \bf S +\bf L\cdot\bf A) \;\; , \label{mmnp}  \\
v_{MM}(nn) & = & -{\alpha\over 4M^2_n}\mu^2_n\left[{2\over 3}F_\delta(r)
  \mbox{\bm$\sigma$}_i\cdot\mbox{\bm$\sigma$}_j+{F_t(r)\over r^3}S_{ij}\right] \;\; . 
 \label{mmnn}
\end{eqnarray}
 
In the above formula $F_\delta, F_t$ and $F_{ls}$ describe the finite size
of the nucleon charge distributions. As $r\rightarrow\infty$, 
$F_\delta\rightarrow 0$ whereas $F_t\rightarrow 1$ and $F_{ls}\rightarrow 1$.
$M_p$ ($M_n$) is the proton (neutron)
mass and $M_{np}$ is the $n-p$ reduced mass. The MM interaction presents
the usual $r^{-3}$ behavior and has an operatorial structure with a
spin-spin, a tensor and a spin-orbit term. In the $n-p$ case, this last term includes 
an asymmetric force (proportional to 
${\bf A}=$(\mbox{\bm$\sigma$}$_i-$\mbox{\bm$\sigma$}$_j)/2$)
which mixes spin-singlet and spin-triplet states. This term is expected to
have a very small effect.

The EM interaction has been studied in the description of bound states
in $A \le 8$ nucleon systems~\cite{A8}. Recently
a detailed analysis of the contribution of the electromagnetic terms
to the $^3$He--$^3$H mass difference has been performed~\cite{nogga}.
A first analysis in three--nucleon scattering has been done by 
Stoks~\cite{stoks} including the MM interaction in Born approximation
at high $L$ values. However, the $T$-matrices used at low $L$ values were
calculated without considering the MM interaction. In this approximate 
treatment of the process the main modifications were obtained in the
$n-d$ vector analyzing powers at forward angles. No modifications were
observed in other observables as the differential cross section and tensor
analyzing powers and in the maximum of $A_y$ and
$iT_{11}$. As a consequence, the conclusion was that the MM interaction does
not improve the theoretical underestimation of the last two observables.
However, disregarding the MM interaction could not be correct in the
description of low partial waves which govern the polarization observables
at low energies. 
In Refs.~\cite{kie2002,hw2002} the MM interaction has been included
in the calculation of $N-d$ scattering observables. However in these analyses
its contribution was limited to
a low number of partial waves. The contribution from waves with high $L$ values
was neglected. In the present paper we will include the MM interaction in
both regimes in order to perform a complete description of the collision
process.

For the case $A=2$, the contribution of the MM interaction to the
scattering amplitude has been extensively studied~\cite{knutson78,stoks90}.
It has been shown that due to its $r^{-3}$ behavior the scattering
amplitude results in a slow convergent series whose leading term can be summed
analytically. A similar analysis can be performed for
$N-d$ scattering. The starting point is the $N-d$ transition matrix
$M$ which can be decomposed as a sum of the Coulomb amplitude $f_c$ plus
a nuclear term, namely
\begin{eqnarray}
M^{SS'}_{\nu\nu'}(\theta)& = &f_c(\theta)\delta_{SS'}\delta_{\nu\nu'}+
{\sqrt{4\pi}\over k}\sum_{L,L',J}\sqrt{2 L+1}(L0S\nu|J\nu)(L'M'S'\nu'|J\nu) 
\nonumber \\
&& \,\exp[i(\sigma_L+\sigma_{L'}-2\sigma_0)]\;
 {}^JT^{SS'}_{LL'} \; Y_{L'M'}(\theta,0) \;\; .
\label{tm}
\end{eqnarray}

This is a $6\times6$ matrix corresponding to the two possible couplings of the 
spin $1$ of the deuteron and the spin $1/2$ of the third particle 
to $S,S'=1/2$ or $3/2$ and their projections $\nu$ and $\nu'$.
The quantum numbers $L,L'$ represent the relative orbital angular momentum between
the deuteron and the third particle and $J$ is the total angular momentum
of the three-nucleon scattering state. 
${}^JT^{SS'}_{LL'}$ are the $T$-matrix elements corresponding to a Hamiltonian
containing nuclear plus Coulomb plus MM interactions and
$\sigma_L$ are the Coulomb phase--shifts. The $n-d$ case
is recovered putting $f_c=\sigma_L=0$. When the MM
interaction is not considered the sums over $L$, $L'$, $J$ converge very fast 
due to the finite range of the nuclear interactions. Typically in the low energy
region ($E_{lab}<50$ MeV) states with $L,L'>10$ can be safely neglected. However,
when the MM interaction is considered, an infinite number of terms contributes
to the construction of the scattering amplitude. In this case the sums on 
$L,L'$ can be divided in two parts. For $L,L' \le L_{max}$ the
$T$--matrix elements correspond to, and are obtained from,
a complete three-body description of the system. 
For $L,L' > L_{max}$ the centrifugal barrier is sufficiently
high to maintain the third particle far from the deuteron and the description
of the state can be performed as a two-body system. 
In general $L_{max}$ can be fixed in such a way that when the collision
proceeds in states with $L,L' > L_{max}$
the nuclear interaction can be safely neglected and
only the Coulomb plus MM potentials contributes to the $N-d$ scattering. 
It is therefore convenient to introduce the MM interaction between a nucleon and
the deuteron as distinct particles.
Its specific form can be obtained summing the MM interaction
between each nucleon of the deuteron and the third nucleon at large
separation distances. Alternatively,
the $N-d$ MM interaction can be obtained directly in one-photon exchange
approximation between a spin-1 and a spin-1/2 particle from
a non-relativistic reduction of the corresponding Feynman diagram. 
Here below the MM $n-d$ and $p-d$ interactions are explicitly
given. The details of the derivation are reserved to the Appendix.
\begin{eqnarray}
v_{MM}(nd)&=&-{\alpha\over r^3}[ {\mu_n\mu_d\over M_n M_d} S_{nd}^I
       +{\mu_n\over 2 M_n M_{nd}}
        ({\bf L}\cdot{\bf S}_{nd}+{\bf L}\cdot{\bf A}_{nd})]\;\; , \label{ndMM} \\
v_{MM}(pd)&=&-{\alpha\over r^3}[ {\mu_p\mu_d\over M_p M_d} S_{pd}^I
       +({\mu_p\over 2 M_p M_{pd}}-{1\over 4 M^2_p})
        ({\bf L}\cdot{\bf S}_{pd}+{\bf L}\cdot{\bf A}_{pd})\nonumber \\
    && +({\mu_d\over 2 M_d M_{pd}}-{1\over 4 M^2_d})
        ({\bf L}\cdot{\bf S}_{pd}-{\bf L}\cdot{\bf A}_{pd})
       -{Q_d\over 2} S^{II}_d]\;\; ,  \label{pdMM} \\
    S^I_{Nd} &=& 3({\bf S}_N\cdot{\hat r})({\bf S}_d\cdot{\hat r})
       -{\bf S}_N\cdot{\bf S}_d,\;\;\;\; N=n,p  \\
    S^{II}_d &=& 3({\bf S}_d\cdot{\hat r})^2 - 2  \;\; ,
\label{ndMMf}
\end{eqnarray}
where $M_d$ is the deuteron
mass, $M_{Nd}$ is the corresponding nucleon-deuteron reduced mass and $\mu_d$,
$Q_d$ are the magnetic and the quadrupole moments of the deuteron, respectively.
Moreover,
${\bf S}_{Nd}={\bf S}_N+{\bf S}_d$ whereas ${\bf A}_{Nd}={\bf S}_N-{\bf S}_d$.
The deuteron-nucleon distance is $r$ and $\hat r$ is the unitary vector giving 
their relative position.

\subsection{$n-d$ case}

Let us first discuss $n-d$ scattering including the MM interaction. For relative 
states verifying $L,L'>L_{max}$ the description proceeds as a two-body
process and the 
$T$--matrix elements corresponding to a $n-d$ state with total
angular momentum $J$, relative angular momentum $L$ and total spin
$S$ are given in Born approximation as 
\begin{equation}
  ^J\!T^{LL'}_{SS'}=-k ({2M_{nd}\over \hbar^2})
  <\Omega_{L'S'J}|v_{MM}(nd)|\Omega_{LSJ}>  \;\; .
\label{born1}
\end{equation}
The relative motion of the $n-d$ system is described by the regular 
free solution of Schr\"odinger equation
\begin{equation}
  \Omega_{LSJ}=j_L(kr)[Y_L(\hat r)\otimes \chi_S]_{JJ_z}  \;\; ,
\label{ond}
\end{equation}
with $k^2=(2M_{nd}/\hbar^2)E_{cm}$, $j_L$ a spherical Bessel function and
$\chi_S$ the total spin function.

The $T$--matrix elements corresponding to the
spin-orbit term of the MM interaction proportional to
${\bf L}\cdot{\bf S}+{\bf L}\cdot{\bf A}$ are
\begin{equation}
  ^J\!T^{LL'}_{SS'}=-k\; C_{so} <\Omega_{L'S'J}|
  {{\bf L}\cdot{\bf S}+{\bf L}\cdot{\bf A}\over r^3}|\Omega_{LSJ}>
  =-k\; C_{so}{\delta_{LL'}\over 2L(L+1)}\;\; ^J\!M^L_{SS'}  \;\; ,
\label{km}
\end{equation}
with
\begin{equation}
  C_{so}=-{\alpha\mu_n\over M_n}\approx 2.932\times 10^{-3}\;\;{\rm fm} 
\end{equation}
and 
\begin{equation}
  ^J\!M^L_{SS'}=(-1)^{L+J+S-S'-1/2}
  \sqrt{6(2S+1)(2S'+1)}
  \left\{ \begin{array}{rrr}\frac{1}{2}&S'&1\\S&\frac{1}{2}&1\end{array} \right \}
  \sqrt{L(L+1)(2L+1)}
  \left\{ \begin{array}{rrr} S'& L & J \\ L & S & 1 \end{array} \right \} \ . 
\label{mnd}
\end{equation}

The $T$-matrix elements of Eq.(\ref{km}) can be
used in Eq.(\ref{tm}) for values of $L,L'>L_{max}$. Moreover, for fixed values
of $L$ the sum over $J$ can be performed analytically using
summation properties of Clebsh-Gordan coefficients. The convergence
of the sum on $L$ is slow enough to prevent a safe truncation of the
series. Therefore, after summing all terms for $L>L_{max}$,
the contribution of the spin-orbit term
to the transition matrix of Eq.(\ref{tm}) results:

\begin{equation}
  M^{SS'}_{\nu\nu'}(so)=\frac{C_{so}}{2}K^{SS'}_{\nu\nu'}
 \left[\frac{\sin\theta}{1-\cos\theta}-
 \sum_{L=1}^{L_{max}}\frac{(2L+1)}{L(L+1)} P^1_L(\cos\theta) \right] \;\; .
\end{equation}
$P^1_L(\cos\theta)$ is a generalized Legendre polynomial and the following
property has been used to derive the above equation
\begin{equation}
 \sum_{L=1}^\infty\frac{(2L+1)}{L(L+1)} P^1_L(\cos\theta)=
 \frac{\sin\theta}{1-\cos\theta} \;\; .
\end{equation}
Moreover
\begin{equation}
 K^{SS'}_{\nu\nu'}= (-1)^{S-S'+\nu+\frac{1}{2}}
 \sqrt{3(2S+1)(2S'+1)}
 \left\{ \begin{array}{rrr} \frac{1}{2}& S'& 1 \\ S & \frac{1}{2} & 1 \end{array} 
 \right \}
 \left(  \begin{array}{rrr} S & S' & 1 \\ \nu & -\nu' & -M \end{array} \right )
 \delta_{|M|,1}\; .
 \label{ksnd}
\end{equation}

As a consequence of the $r^{-3}$ behavior of the MM spin-orbit interaction
a term proportional to $\sin\theta/(1-\cos\theta)$ appears in the transition
matrix. This term produces a divergence in the differential cross section at
extreme forward angles and a pronounced dip structure in the vector analyzing 
powers. 

A similar analysis can be done for the term proportional to the tensor
operator in the $n-d$ MM interaction. The corresponding $T$--matrix elements are
\begin{equation}
  ^J\!T^{LL'}_{SS'}=-k\; C_t <\Omega_{L'S'J}|
  {S_{nd}^I \over r^3}|\Omega_{LSJ}>
  =-k\; C_t\; I_{LL'}\;\; ^J\!M^{LL'}_{SS'} \;\; ,
\label{ktm}
\end{equation}
with
\begin{equation}
  C_t=-{\alpha\mu_n\mu_d\over M_n+ M_d}\approx  1.675 \times 10^{-3} \;
  \; {\rm fm}  \;\;\ .
\end{equation}
The angular-spin and radial matrices are
\begin{eqnarray}
  ^JM^{LL'}_{SS'}&=&(-1)^{L+L'+J+S'}
  \left\{ \begin{array}{rrr} \frac{1}{2}&1& S' \\ \frac{1}{2}&1&S \\
                             1&1&2 \end{array} \right \}  \nonumber \\
  &\times&
  \sqrt{30(2L+1)(2L'+1)(2S+1)(2S'+1)}
  \left\{ \begin{array}{rrr} L'& S' & J \\ S & L & 2 \end{array} \right \}
  \left( \begin{array}{rrr} L & 2 & L' \\ 0 & 0 & 0 \end{array} 
  \right )
\end{eqnarray}
and
\begin{equation}
  I_{LL'}=
  \left\{ \begin{array}{l} \delta_{LL'}\over 2L(L+1) \\ 
                           \delta_{L+2,L'}\over 6(L+1)(L+2) \\
  \delta_{L-2,L'}\over 6(L'+1)(L'+2) \end{array} \right. 
\end{equation}

Again for fixed values of $L$ and $L'$ the sum over $J$ in Eq.(\ref{tm}) 
can be performed analytically
and the contribution to the transition matrix is
\begin{eqnarray}
  M^{SS'}_{\nu\nu'}(t)&=&-\sqrt{4\pi}C_t \sqrt{30(2S+1)(2S'+1)}
  \left\{ \begin{array}{rrr} \frac{1}{2}&1& S' \\ \frac{1}{2}&1&S \\
                             1&1&2 \end{array} \right \}
(-1)^{S'-\nu} \left( \begin{array}{rrr} S & S' & 2 \\ \nu & -\nu'& -M \end{array}  
              \right) \nonumber \\
 &\times&\sum_{L,L'>L_{max}} (2L+1)\sqrt{2L'+1} I_{LL'}
  \left( \begin{array}{rrr} L & L' & 2 \\ 0 & 0 & 0 \end{array} \right) 
  \left( \begin{array}{rrr} L' & L & 2 \\ -M & 0 & M \end{array} 
  \right ) Y_{L'M}(\theta,0) \;\; .
\end{eqnarray}

Three different sums can be constructed corresponding to $|M|=0,1,2$
that can be summed numerically term by term. The convergence
of the series is rather fast and a few tens of terms are sufficient.

In conclusion, the $n-d$ transition matrix including the nuclear plus the
MM interaction can be constructed as a sum of three terms
\begin{eqnarray}
M^{SS'}_{\nu\nu'}(\theta)&=&{\sqrt{4\pi}\over k}\sum_{L,L'}^{L_{max}}
\sum_J\sqrt{2L+1}(L0S\nu|J\nu)(L'M'S'\nu'|J\nu) 
 \;{}^JT^{SS'}_{LL'} \; Y_{L'M'}(\theta,0) \nonumber \\
 &&+M^{SS'}_{\nu\nu'}(so) +M^{SS'}_{\nu\nu'}(t) \;\;\ .
\label{tmnd}
\end{eqnarray}

When the MM interaction is neglected only the first term contributes 
to the transition matrix. When the MM interaction
is included, the $T$--matrix elements for $L,L'\le L_{max}$
are different from the previous case. In addition the last two terms 
in Eq.(\ref{tmnd}) have to be included. We stress the fact that
the value of $L_{max}$ can be taken in such a way that for $L,L'>L_{max}$ 
the nuclear interaction gives a negligible contribution to the scattering process
and the interaction between the incident particle and the target is only
electromagnetic. Typical values for $L_{max}$ are discussed in Sec.IV.

\subsection{$p-d$ case}

As for the $n-d$ case, the $T$--matrix elements corresponding to a 
two-body description of the $p-d$ system with total
angular momentum $J$, relative angular momentum $L$ and total spin
$S$, are given in Born approximation
\begin{equation}
  ^J\!T^{LL'}_{SS'}=-k ({2M_{pd}\over \hbar^2})
  <\Omega_{L'S'J}|v_{MM}(pd)|\Omega_{LSJ}>  \;\; .
\label{born2}
\end{equation}
Here the relative motion of the $p-d$ system is described by 
\begin{equation}
  \Omega_{LSJ}=F_L(\eta,kr)[Y_L(\hat r)\otimes \chi_S]_{JJ_z}
\label{opd}
\end{equation}
with $k^2=(2M_{pd}/\hbar^2)E_{cm}$, $F_L$ a regular Coulomb function
and $\eta$ the usual Coulomb parameter.

Let first consider the
spin-orbit terms of the MM interaction in Eq.(\ref{pdMM}) proportional to
$({\bf L}\cdot{\bf S}+{\bf L}\cdot{\bf A})$ and
$({\bf L}\cdot{\bf S}-{\bf L}\cdot{\bf A})$. The following matrix elements
entering in the calculation of the $T$--matrix are defined
\begin{equation}
<\Omega_{L'S'J}|{{\bf L}\cdot{\bf S}\pm {\bf L}\cdot{\bf A}\over r^3}|\Omega_{LSJ}>
=I_{LL'}\;\delta_{LL'}\; ^J\!M^L_{SS'}(\pm)
\label{pdkm}
\end{equation}
with~\cite{bc55}
\begin{equation}
I_{LL}=\frac{1}{2L(L+1)}+\frac{1}{2L(L+1)(2L+1)}\left[\eta\pi+1+\eta\pi
{\rm coth}\eta\pi-2\eta^2\sum_{p=0}^L\frac{1}{p^2+\eta^2}\right] \;\; .
\label{pdll}
\end{equation}

In Eq.(\ref{pdkm}) the angular-spin matrix $\;^J\!M^L_{SS'}(+)$ is equal
to the matrix $^J\!M^L_{SS'}$ defined in Eq.(\ref{mnd}), whereas
\begin{equation}
  ^J\!M^L_{SS'}(-)=(-1)^{J+L-1/2}
  \sqrt{24(2S+1)(2S'+1)}
  \left\{ \begin{array}{ccc} 1 & S' & \frac{1}{2} \\ S & 1 & 1 \end{array} \right \}
  \sqrt{L(L+1)(2L+1)}
\left\{ \begin{array}{ccc} S' & L' & J \\ L & S & 1 \end{array} \right \} .
\end{equation}

Following Ref.~\cite{knutson78} we isolate the first term of $I_{LL}$ and proceed
toward a summation of the related amplitude as we have done for the $n-d$ case.
The corresponding contribution to the transition matrix of Eq.(\ref{tm})
for $L>L_{max}$ results
\begin{eqnarray}
   M^{SS'}_{\nu\nu'}(so)& =& \frac{1}{2}[C^+_{so} K^{SS'}_{\nu\nu'}(+)
        +C^-_{so} K^{SS'}_{\nu\nu'}(-)]  \nonumber \\ &\times &
 \left[\frac{\cos\theta+2{\rm e}^{-i\eta{\rm ln}(\frac{1-\cos\theta}{2})}-1}
 {\sin\theta}-
 \sum_{L=1}^{L_{max}}\frac{(2L+1)}{L(L+1)} {\rm e}^{2i(\sigma_L-\sigma_0)}
   P^1_L(\cos\theta) \right] \;\; .
\label{fso}
\end{eqnarray}
To get this final form we have used the following analytical
summation of the series~\cite{knutsonp}

\begin{equation}
 \sum_{L=1}^\infty\frac{(2L+1)}{L(L+1)} {\rm e}^{2i\sigma_L} P^1_L(\cos\theta)= 
 \frac{{\rm e}^{2i\sigma_0}}{\sin\theta}
 [\cos\theta+2{\rm e}^{-i\eta{\rm ln}(\frac{1-\cos\theta}{2})}-1] \;\; ,
\end{equation}
which can be obtained from the series of the Coulomb amplitude
\begin{equation}
f_c(\theta)=
 \sum_{L=0}^\infty(2L+1)({\rm e}^{2i\sigma_L}-1) P_L(\cos\theta)= 
 -2i\eta\frac{{\rm e}^{2i\sigma_0}}{1-\cos\theta}
 {\rm e}^{-i\eta{\rm ln}(\frac{1-\cos\theta}{2})} \;\; ,
\end{equation}
using the recurrence relations of the Legendre polynomials and the
following relation of the Coulomb phase-shifts
\begin{equation}
 {\rm e}^{2i(\sigma_L-\sigma_{L-1})} (L-i\eta)=L+i\eta \;\; .
\end{equation}

In Eq.(\ref{fso}) $K^{SS'}_{\nu\nu'}(+)=K^{SS'}_{\nu\nu'}$ defined in Eq.(\ref{ksnd})
and 
\begin{equation}
 K^{SS'}_{\nu\nu'}(-)= (-1)^{\nu+\frac{1}{2}}
 \sqrt{12(2S+1)(2S'+1)}
 \left\{ \begin{array}{rrr} 1& S'& \frac{1}{2} \\ S & 1 & 1 \end{array} 
 \right \}
 \left(  \begin{array}{rrr} S & S' & 1 \\ \nu & -\nu' & -M \end{array} \right )
 \delta_{|M|,1} \;\; .
 \label{kspd}
\end{equation}

Moreover
\begin{eqnarray}
  C_{so}^+=-\alpha M_{pd}\left(\frac{\mu_p}{M_pM_{pd}}-\frac{1}{2M^2_p}
 \right)\approx -3.775\times 10^{-3}\;\;{\rm fm}\;\; ,    \\
  C_{so}^-=-\alpha M_{pd}\left(\frac{\mu_d}{M_dM_{pd}}-\frac{1}{2M^2_d}
 \right)\approx -5.936\times 10^{-4}\;\;{\rm fm} \;\; .
\end{eqnarray}
The term proportional to $C_{so}^-$ is much smaller due to the small
magnetic moment of the deuteron. The same happens to the term proportional
to $Q_d$ in Eq.(\ref{pdMM}) due to the small quadrupole moment of the
deuteron and will not be discussed here. The analysis of the term
proportional to the tensor operator in the MM interaction proceeds
similarly to that one performed in the $n-d$ case, taking care that now
the radial integral $I_{LL}$ is given by Eq.(\ref{pdll}) and
$I_{L,L+2}=\frac{1}{6}|L+1+i\eta|^{-1}|L+2+i\eta|^{-1}$~\cite{bc55}.
In conclusion the transition matrix can be constructed as a sum of different
contributions
\begin{eqnarray}
M^{SS'}_{\nu\nu'}(\theta)& = &f_c(\theta)\delta_{SS'}\delta_{\nu\nu'}+
{\sqrt{4\pi}\over k}\sum_{LL'}^{L_{max}}\sum_{J}\sqrt{2 L+1}
(L0S\nu|J\nu)(L'M'S'\nu'|J\nu) 
\nonumber \\
&\times& \,\exp[i(\sigma_L+\sigma_{L'}-2\sigma_0)]\;
 {}^JT^{SS'}_{LL'} \; Y_{L'M'}(\theta,0)
 + M^{SS'}_{\nu\nu'}(so)+B^{SS'}_{\nu\nu'} \;\; ,
\label{tmpd}
\end{eqnarray}
where $M^{SS'}_{\nu\nu'}(so)$ is defined in Eq.(\ref{fso}) and
$B^{SS'}_{\nu\nu'}$ includes the contribution of the remaining terms 
in Eq.(\ref{pdll}) and those coming from the tensor operator. 
The $B^{SS'}_{\nu\nu'}$ matrix elements can be evaluated summing the corresponding
series numerically for $L,L'>L_{max}$ until convergence is reached.

\section{The $3N$ and $N-d$ $T$-matrices in Born approximation}

The calculations of the observables in $N-d$ scattering can be obtained
from the transition matrices of Eqs.(\ref{tmnd}) and (\ref{tmpd}). Accordingly,
after a partial wave decomposition, it is necessary to calculate
the three-nucleon $T$--matrices for states with total angular momentum $J$
in which the deuteron and the incident nucleon are in
relative motion in the regime $L\le L_{max}$. As discussed before, states
having $L>L_{max}$ are described as a two-body process. 
Therefore it is appropriate to make a link
between the two regimes and show in which manner the three-nucleon
$T$--matrix smoothly tends to a two-body $T$--matrix as $L$ increases.

The KVP in its complex form establishes that 
the $T$-matrix elements are functionals of the 
three-nucleon scattering state
\begin{equation}
[{}^J\!{T}^{SS'}_{LL'}]= {}^J\!{T}^{SS'}_{LL'}-\frac{M}{2\sqrt{3}\hbar^2}
\langle\Psi^-_{LSJ}|H-E|\Psi^+_{L'S'J}\rangle \ .
\label{ckohn}
\end{equation}
The stationarity of this
functional with respect to the trial parameters in the
three--nucleon scattering state $\Psi^+_{LSJ}$ is required to obtain
the $T$--matrix first order solution. The second order estimate is obtained
after replacing the first order solution in Eq.(\ref{ckohn}). 
In this formalism~\cite{kohn} the continuum state is usually written
as a sum of
three Faddeev-like amplitudes, each of which consists of two terms:
\begin{equation}
\Psi^+_{LSJ}=\sum_{i=1,3}\left[ \Psi_C(\x_i,\y_i)+\Omega^+_{LSJ}(\x_i,\y_i)
             \right] \;\;\; ,
\end{equation}
here ${\bf x}_i,{\bf y}_i$ are the Jacobi coordinates corresponding
to the $i$--th permutation of the particles indices $1,2,3$.
The first term, $\Psi_C$,
describes the system when the three--nucleons are close to each other. 
For large interparticle separations and energies below the
deuteron breakup threshold it goes to zero, whereas for higher energies it must
reproduce a three outgoing particle state. 
The second term, $\Omega^+_{LSJ}$, describes the asymptotic configuration 
of a deuteron far from the third nucleon and explicitly it is:
\begin{equation}
\Omega^+_{LSJ}(\x_i,\y_i) =  \Omega^0_{LSJ}(\x_i,\y_i)+
 \sum_{L'S'}{}^J\!T^{SS'}_{LL'}\Omega^1_{L'S'J}(\x_i,\y_i)  \ ,
\end{equation}
where
\begin{eqnarray}
\Omega^0_{LSJ}(\x_i,\y_i) &=& \Omega^R_{LSJ}(\x_i,\y_i) \\
\Omega^1_{LSJ}(\x_i,\y_i) &=& \Omega^R_{LSJ}(\x_i,\y_i)-
                            i\Omega^I_{LSJ}(\x_i,\y_i)  \ .
\end{eqnarray}
Besides a factor $\sqrt{k}$,
$\Omega^R_{LSJ}$ is the function $\Omega_{LSJ}$ given in Eqs.(\ref{ond}) and
(\ref{opd}) for
the $n-d$ and $p-d$ system respectively, in which $\chi_S$ represents the deuteron
wave function of spin $1$ coupled with the spin $\frac{1}{2}$ of the third 
nucleon to total spin $S$. In $\Omega^I_{LSJ}$
the regular relative function $j_L$ or $F_L$ is replaced by the corresponding
irregular solution $\eta_L$ or $G_L$ regularized at the origin\cite{phh1}.
The normalization of the asymptotic states verifies
\begin{equation}
\frac{M}{2\sqrt{3}\hbar^2}\sum_{i,j}
[\langle\Omega^R_{LSJ}(\x_i,\y_i)|H-E|\Omega^I_{L'S'J}(\x_j,\y_j)\rangle 
-\langle\Omega^I_{LSJ}(\x_i,\y_i)|H-E|\Omega^R_{L'S'J}(\x_j,\y_j)\rangle] =1 \;\; ,
\label{norm}
\end{equation}
$M$ being the nucleon mass.
To be noticed that in the three-nucleon process the energy in
the center of mass reference frame is
\begin{equation}
E={4M k^2\over 3 \hbar^2}+E_d = {2M_{Nd} k^2 \over \hbar^2} +E_d
\end{equation}
with $E_d$ the deuteron ground state energy.
Moreover,
the factor $1/(2\sqrt{3})$ in Eq.(\ref{norm}) is related to the definitions of the 
Jacobi coordinates in terms of the particle coordinates:
\begin{eqnarray}
\x_i &=& {\bf r_j}-{\bf r_k} \nonumber \\
\y_i &=& \frac{2}{\sqrt{3}}({\bf r_k}-\frac{{\bf r_i}+{\bf r_j}}{2}) \ .
\end{eqnarray}

The Born approximation of the $T$-matrix is obtained from Eq.(\ref{ckohn})
replacing the wave function $\Psi$ by the regular function $\Omega^0$
and putting the first order $T$-matrix equal to zero:
\begin{equation}
[{}^J\!{T}^{SS'}_{LL'}]_B=-\frac{M}{2\sqrt{3}\hbar^2}\sum_{i,j}
\langle\Omega^0_{LSJ}(\x_i,\y_i)|H-E|\Omega^0_{L'S'J}(\x_j,\y_j)\rangle \ .
\label{born}
\end{equation}

For a given energy a certain value $L_B$ exists such that for
$L,L'\ge L_B$ the differences between the $T$--matrix elements obtained from
a complete solution of the three-nucleon problem or from its Born
approximation are extremely small. Increasing further the values of $L$
and $L'$ we arrive to the regime $L,L'>L_{max}$ in which the contribution of
the $NN$ nuclear potential can be neglected.
Let us consider $\x_3,\y_3$ the Jacobi coordinates corresponding to the
asymptotic configuration in which nucleons ($1,2$) form the deuteron 
and nucleon $3$ is the incident particle. 
The relative coordinate between the third nucleon and the center of mass
of the deuteron is ${\bf r}_{Nd}=(\sqrt{3}/2)\y_3$. 
Starting from the above Born approximation for the $T$-matrix, the 
following relations are verified for $L,L'>L_{max}$:
\begin{eqnarray}
[{}^J{T}^{SS'}_{LL'}]_B& =& -\frac{M}{2\sqrt{3}\hbar^2}\sum_{i,j}
\langle\Omega^0_{LSJ}(\x_i,\y_i)|H-E|\Omega^0_{L'S'J}(\x_j,\y_j)\rangle \\
& =& -3\frac{M}{2\sqrt{3}\hbar^2}\sum_i
\langle\Omega^0_{LSJ}(\x_i,\y_i)|V(1,3)+V(2,3)|\Omega^0_{L'S'J}(\x_3,\y_3)\rangle \\
& \approx& -3\frac{M}{2\sqrt{3}\hbar^2}
\langle\Omega^0_{LSJ}(\x_3,\y_3)|V(1,3)+V(2,3)|\Omega^0_{L'S'J}(\x_3,\y_3)\rangle \\
& \approx& -3\frac{M}{2\sqrt{3}\hbar^2}
\langle\Omega^0_{LSJ}(\x_3,\y_3)|v_{MM}(Nd)|\Omega^0_{L'S'J}(\x_3,\y_3)\rangle  \\
& = & -2k\frac{M_{Nd}}{\hbar^2}
\langle\Omega_{LSJ}|v_{MM}(Nd)|\Omega_{L'S'J})\rangle  \ . 
\end{eqnarray}
The equivalence between the second and third row is in general verified 
for $L,L'>L_B$. On the other hand, the equivalence between the third and 
fourth row is verified for $L,L'>L_{max}$. In fact, $L_{max}$ can be fixed as
the $L$ value at which these two rows start to be approximately equal. 
Finally, in the last step the asymptotic three-nucleon function $\Omega^0$ 
has been replaced
by the two-body function $\Omega$ of Eq.(\ref{ond}) once the integration
over $\x_3$ and the change of variables $\y_3\rightarrow{\bf r}_{Nd}$ has been
performed. In conclusion, the above approximate
equalities show the relation between the three--nucleon $T$--matrix
of Eq.(\ref{born}) and the two-body $T$--matrices of Eqs.(\ref{born1}) and
(\ref{born2}) for high $L$ values.

\section{$N-d$ observables including Coulomb plus MM interactions}

Elastic observables for $N-d$ scattering can be calculated using
the transition matrices of Eqs.(\ref{tmnd}) and (\ref{tmpd}) using trace
operations~\cite{seyler}. The calculations presented here have
been performed using the KVP after an expansion of the three-nucleon
scattering wave function in terms of the pair correlated hyperspherical
harmonic (PHH) basis~\cite{phh1,phh2}.
As $NN$ interaction we have used the nuclear part of the AV18
potential plus the Coulomb and MM interactions defined in 
Eqs.(\ref{mmpp})-(\ref{mmnn}). The asymmetric force ${\bf L}\cdot{\bf A}$
in the $v_{MM}(np)$ interaction has been included as well as the $p-p$
Darwin-Foldy and the $n-p$ short-range Coulomb terms.

At energies below the deuteron breakup threshold
the contribution of the MM interaction is
expected to be appreciable. Recently the $n-d$ 
analyzing power $A_y$ has been measured at $E_{lab}=1.2$ and $1.9$ 
MeV~\cite{werner2002}. At these very low energies the nuclear part
of the transition matrix (first term of Eq.(\ref{tmnd})) converges
already for $L_{max}=3$. The corresponding theoretical curves 
obtained using the AV18 potential, and neglecting the MM interaction,
are showed in Fig.1 (solid line). As it can be seen, the observable is not
reproduced by a large amount which is a common feature of all modern $NN$ forces. 
When the MM interaction is taken into account up to $L_{max}=3$,
the analyzing powers are given by the dashed curves. There is
a very small influence of the MM interaction in the peak of $A_y$ 
with the tendency of slightly flattening the observable. 
However, this is an incomplete calculation since the inclusion of the
MM interaction requires an infinite number of partial waves in the 
calculation of the transition matrix. When all three terms of
Eq.(\ref{tmnd}) are considered the observables are given by the 
dashed-dotted curves. It is interesting to notice the forward-angle
dip structure which already appears in $n-p$ scattering~\cite{stoks90}.
Only after summing the series up to $\infty$ this particular behavior can be
reproduced. We can conclude that the MM interaction
produces a pronounced modification of $A_y$ at forward angles but has a
very small effect around the peak. 

In order to show the importance of the MM moment interaction in the
calculations of $A_y$ as the
energy increases, in Fig.2 the results at $E_{lab}=6.5$ MeV are given.
At this particular energy $A_y$ has been measured in an extended angular
range including forward angles~\cite{werner86}. The solid line corresponds
to a standard AV18 calculation neglecting the MM interaction and
including partial waves up to $L_{max}=8$. The dashed-dotted line corresponds
to a calculation using the AV18+MM potential and considering the complete series.
We can observe that the effect of the MM interaction on the peak is
practically negligible. Conversely, it is of great importance at
forward angles in order to describe the zero crossing.

Besides the neutron analyzing power and the deuteron analyzing power which 
present similar characteristics, other elastic $n-d$ observables as the tensor
analyzing powers suffer only
minor modifications when the MM interaction is included. The differences
are of the order of $1$\% or less and they are not presented here.
However when comparisons with precise experimental data are performed these
differences could be relevant and the MM interaction should be taking
into account.

For $p-d$ scattering high precision data exist at low 
energies~\cite{brune,wood,knutson,sagara} for differential cross section
and vector and tensor analyzing powers. Detailed comparisons to these
data has been performed in Refs.~\cite{puzzle,brune,wood,kievwood}
using AV18 with and without the inclusion of three--nucleon forces. In those
studies the Coulomb interaction was included whereas the MM interaction
was not. In order to evaluate the effects of the MM interaction on the
vector analyzing powers in presence of the Coulomb field, in Fig.3 the results 
of the calculations at $E_p=1$ and $3$ MeV are shown.
Three different calculations have been performed at both energies. The solid
line corresponds to the AV18 prediction neglecting the MM interaction. 
Accordingly, the transition matrix has been calculated
with the first two terms of Eq.(\ref{tmpd}). The partial-wave series of the
second term has been summed
up to $L_{max}=4$ ($E_p=1$ MeV) and $L_{max}=6$ ($E_p=3$ MeV). 
The dashed line corresponds to the same calculation as before but the
$T$-matrix elements has been calculated using the AV18+MM potential.
The dashed-dotted line corresponds to the complete calculation
including also the last two terms of Eq.(\ref{tmpd}). We see that the major effect
of the MM interaction is obtained around the peak and is appreciable
at both energies. There is also an improvement in the description of
the observable at forward angles, in particular for $iT_{11}$ at $E_p=3$ MeV. 
The observed modifications are due to the
interference between the Coulomb and the nuclear plus the MM interaction
and not to higher order terms, as in the $n-d$ case, since, except for
$A_y$ at $E_p=1$ MeV, the dashed and dashed-dotted line practically overlap. 
In fact, high order terms are dominated by the Coulomb interaction
and the MM interaction gives a very small contribution.

As the energy increases, the effect of the MM interaction on $A_y$ and $iT_{11}$
diminishes as it is shown in Fig.4 at $E_p=5$ and $10$ MeV. 
Here the AV18 prediction (solid line) has to be
compared to the AV18+MM prediction (dashed line) calculated using the first
two terms of Eq.(\ref{tmpd}) with $L_{max}=8$. When the
last two terms of Eq.(\ref{tmpd}) are also included, the results are extremely
close to the previous ones. As for the $n-d$, the tensor analyzing
powers present very small modifications when the MM interaction is taking
into account and are not presented here.

The MM interaction has different effects in $n-d$ or $p-d$
vector analyzing powers. One reason is the different sign between the neutron 
and proton magnetic moment. Another reason is the interference with the 
Coulomb field. However the MM interaction does not help for a better description
of the neutron $A_y$. On the contrary there is an appreciable improvement in the
proton $A_y$ as well as in $iT_{11}$, in particular at very low energies. 
Hence we can examine the differences between the experimental data and the 
theory at the peak in order to see if the
inclusion of the MM interaction helps to clarify a different behavior observed 
for $n-d$ and $p-d$. In Fig.5 the relative difference 
$[A_y(exp)-A_y(th)]/A_y(exp)$ at the peak for $n-d$ and for $p-d$ scattering
is shown. In this
last case both, the AV18 and AV18+MM results have been reported. For $n-d$ both
results are extremely close at the peak, so the difference does not depend
on which calculation (AV18 or AV18+MM) is considered.
Without the inclusion of the MM interaction
the underestimation of the proton $A_y$ is much more pronounced than the
neutron $A_y$. When the MM interaction is considered the difference between
theory and experiment for both, $p-d$ and $n-d$ scattering
are of similar size, around $25$\%, for all the energy values below $16$ MeV. 
Above $16$ MeV the differences at the peak between theory and experiment diminish.
As shown in Fig.5, at $18$ MeV the difference is around $20$\%.
In Fig.6 the deuteron analyzing power $iT_{11}$ is examined.
The relative difference $[iT_{11}(exp)-iT_{11}(th)]/iT_{11}(exp)$ is shown
at the peak for $p-d$ scattering (there is no data for the $n-d$ case) 
using AV18 and AV18+MM. Besides the first point
at $E_{lab}=0.650$ MeV which corresponds to a very small value of 
$iT_{11}$~\cite{brune}, the underestimation of the observable oscillates 
around $24$\%, very close to the $A_y$ case.

Finally we wish to discuss the importance of the
Coulomb effects as the energy increases.
In fact, up to $E_{lab}=30$ MeV we can observe appreciable
differences in the description of $n-d$ and $p-d$ elastic scattering
that however tend to diminish~\cite{kvr2001}.
Experimental data are not always conclusive since experiments with neutrons
have larger uncertainties than those performed with protons. 
On the other hand, $n-d$ calculations have been often used to describe
$p-d$ scattering, in particular
at high energies~\cite{witala1}. In order to clarify this approximation,
in Figs.7--8 $n-d$ and $p-d$ calculations at $E_{lab}=65$ MeV are compared. 
To make contact with the
results given in Ref.~\cite{witala1} in which $n-d$ scattering
has been analyzed at this particular energy,
we have consider also the Urbana IX (UR) three--nucleon 
interaction~\cite{urbana}.
In Fig.7 the
differential cross section and $A_y$ are shown. Three curves are displayed
corresponding to $p-d$ AV18 (solid line), $n-d$ AV18 (dashed line) and
$p-d$ AV18+UR (dotted line) and compared to the experimental data.
In Fig.8 the same calculations are shown
for $iT_{11}$ and the three tensor analyzing powers $T_{20},T_{21}$ and
$T_{22}$. As expected,
Coulomb effects are small at this energy. We can observe
appreciable Coulomb effects only in $T_{21}$ whereas three-nucleon interaction
effects are found in the minimum of the differential cross
section and in $T_{21}$ and $T_{22}$ as well. These results justify to
some extent the description of $p-d$ data using $n-d$ calculations
at intermediate energies, however with some caution in the description
of particular observables.

\section{conclusions}

The MM interaction has been included in the description of $N-d$ scattering
at low energies. Though its strength is small compared to the nuclear
interaction, it has a very long tail which behaves as $1/r^{3}$.
As a consequence, the construction of the
scattering amplitude necessitates an infinite number of partial waves. 
Analytical summations of the corresponding $p-d$ and $n-d$ series
have been given following previous works for $NN$ scattering. Accordingly,
the $6\times 6$ transition matrix $M$ has been written as a sum of the
standard Coulomb amplitude plus the MM amplitude and a finite series of
$T$--matrices. These matrices have been calculated from a complete
three--body description of the process with a Hamiltonian including the
nuclear plus Coulomb plus MM interaction. For high $L$ values,
the MM amplitude has been calculated as a two-body process. To this aim
the MM interaction between a nucleon and the deuteron as distinct particles
has been derived.

Different $3N$ elastic observables have been calculated and compared to
previous calculations in which the MM interaction was neglected.
The main effect has been observed in the vector analyzing powers. However
the modifications produced by the MM interaction do not
improve the description of the neutron $A_y$ around the peak. Conversely,
there is an appreciable improvement in the proton $A_y$ and in $iT_{11}$, in
particular at low energies. Due
to the different effect that the MM interaction produces in $n-d$ and $p-d$ 
scattering, the relative difference between the calculated and the measured $A_y$ 
at the peak results largely charge independent and approximately
constant below $16$ MeV. The underestimation is about $25$\%.
Above this energy the difference starts to diminish. At $E_{lab}=18$ MeV
it has been reduced to $20$\% and above $30$ MeV there is a much better description 
of $A_y$ and $iT_{11}$. This is shown by the calculations performed
at $E_{lab}=65$ MeV. Furthermore, we have shown that at this energy
Coulomb effects are not important. Only $T_{21}$ still shows some
sensitivity. 

The main aim of this work is to describe the three--nucleon continuum 
using the same $v^{EM}(NN)$ used in the description of the $2N$ scattering
states. In the past the MM interaction has been systematically 
neglected in the calculation of $3N$ scattering observables with few exceptions.
Here we show how to include it and which terms are important. 
From the present analysis it can be concluded
that the approximate treatment of Ref.~\cite{stoks}
is justified for $n-d$ scattering but not for the $p-d$ case. In fact, in the
calculation of the $n-d$ $A_y$ the symmetric spin--orbit term
in $v_{MM}(np)$ tends to depress the observable at the peak whereas
the asymmetric term almost cancel this effect. Therefore the transition
matrix of Eq.(\ref{tmnd}) can be constructed with the MM amplitudes
$M^{SS'}_{\nu\nu'}(so)$ and $M^{SS'}_{\nu\nu'}(t)$ but neglecting the
MM interaction in the calculation of the $T$-matrix elements
$^J\!T^{SS'}_{LL'}$ for $L,L'<L_{max}$. In addition,
the amplitude $M^{SS'}_{\nu\nu'}(t)$
gives an extremely low contribution and can be neglected too.
In the $p-d$ case the interference between the
Coulomb, MM and nuclear interactions does not allow for the omission
of the MM interaction in the calculation of the $T$--matrix elements.
Otherwise the improvement at low energies on the peak of $A_y$ and $iT_{11}$
is lost. However, in the construction of the transition matrix
the last two terms in Eq.(\ref{tmpd}) give very small contributions and,
except at extremely low energies, can be omitted.

Other small terms in the $v^{EM}(pp)$ interaction as the two-photon Coulomb and
vacuum polarization interactions have been neglected in the present analysis.
These terms have improved the description of $p-p$ scattering at low
energies and, therefore, their inclusion in the description of
$p-d$ scattering is of interest. The analysis of these terms as well
as the study of the MM interaction in $p-^3$He scattering is at
present underway.

\acknowledgments
The authors wish to thank S. Rosati for useful discussions and
a critical reading of the manuscript.

\appendix
\section{}
\label{sec:app}

In this appendix we briefly outline the derivation of 
the $N-d$ MM interactions of Eqs.~(\ref{ndMM}) and~(\ref{ndMMf}).

We consider two particles, the first one with spin 1/2, mass, charge and 
magnetic moment  $M_1$, $Z_1$ and $\mu_1$, respectively, 
the second one with spin 1, mass, charge, magnetic moment and quadrupole 
moment $M_2$, $Z_2$, $\mu_2$ and $Q_2$, respectively.
The magnetic and the quadrupole moments are given in n.m. and fm$^2$, 
respectively.
The non-relativistic reduction of the covariant current for 
the point-like spin-1/2 particle gives for the charge $\rho$ 
and current ${\bf j}$ operators in $r$-space~\cite{Car98}:
\begin{eqnarray}
\rho_1({\bf q})&=&Z_1 {\rm e}^{i{\bf q}\cdot{\bf r}_1} 
        -i\,\frac{2\mu_1 -Z_1}{2 M_1^2}
        {\bf q}\cdot({\bf S}_1\times{\bf p}_1)
             {\rm e}^{i{\bf q}\cdot{\bf r}_1} \  , \nonumber \\
{\bf j}_1({\bf q})&=&
             \frac{Z_1}{2 M_1}\{{\bf p}_1,{\rm e}^{i{\bf q}\cdot{\bf r}_1}\}
            -i\,\frac{\mu_1}{M_1} ({\bf q}\times{\bf S}_1)
             {\rm e}^{i{\bf q}\cdot{\bf r}_1} \  ,
\label{eq:rj1}
\end{eqnarray}
where ${\bf q}$ is the three-momentum transferred to the particle, 
${\bf p}_1$ and ${\bf S}_1$ are the momentum and spin operators, respectively,
and $\{\cdots,\cdots\}$ denotes the anticommutator. We have here 
neglected the Darwin-Foldy relativistic correction.

The covariant current operator for a spin-1 particle 
is written as~\cite{Arn80}
\begin{eqnarray}
j^\mu&=&-\frac{1}{\sqrt{4 E\,E'}}\{ [G_1(Q^2)(\epsilon'^*\cdot\epsilon)
        -\frac{G_3(Q^2)}{2M_2^2}(\epsilon'^*\cdot q)(\epsilon\cdot q) ]P^\mu
\nonumber \\
     && + G_2(Q^2) [\epsilon^\mu(\epsilon'^*\cdot q)-
                    \epsilon'^{*\mu}(\epsilon\cdot q) ] \} \ ,
\label{eq:j2mu}
\end{eqnarray}
where $E$, $E'$ are the initial and final energies, 
$\epsilon^\mu\equiv\epsilon(\lambda,p)^\mu$ and 
$\epsilon'^\mu\equiv\epsilon(\lambda',p')^\mu$ are the four-vector spin-1 
initial and final polarizations, $q^\mu=p'^\mu-p^\mu$, 
$P^\mu=p'^\mu+p^\mu$ and $Q^2=-q\cdot q$. 
The three form factors $G_1(Q^2)$, $G_2(Q^2)$ and 
$G_3(Q^2)$ are related to the charge, magnetic and quadrupole form 
factors as~\cite{Car98}
\begin{eqnarray}
G_C(Q^2)&=&G_1(Q^2)+\frac{2}{3}\,\eta\, G_Q(Q^2) \ , \nonumber \\
G_Q(Q^2)&=&G_1(Q^2)-G_2(Q^2)+(1+\eta)G_3(Q^2) \ , \nonumber \\
G_M(Q^2)&=&G_2(Q^2) \ . \label{eq:ff2}
\end{eqnarray}
Here $\eta=Q^2/(4M_2^2)$, $G_C(0)=Z_2$, $G_M(0)=(M_2/M)\, \mu_2$ and 
$G_Q(0)=M_2^2\, Q_2$, $M$ being the nucleon mass.

To perform the non-relativistic reduction of Eq.~(\ref{eq:j2mu}), 
the following relations are used:
\begin{equation}
\epsilon(\lambda,p)^\mu=[\frac{\hat{\bf e}(\lambda)\cdot{\bf p}}{M_2}\, , 
              \hat{\bf e}(\lambda) + 
                     \frac{{\bf p}\,(\hat{\bf e}(\lambda)\cdot {\bf p})}
                                    {M_2(E+M_2)} ] \ ,
\label{eq:eps}
\end{equation}
with $\hat{\bf e}(\pm 1)=\mp\frac{1}{\sqrt{2}}(1,\pm i,0)$, 
$\hat{\bf e} (0)= (0,0,1)$, and 
\begin{equation}
\hat{\bf e}(\lambda ')^*_\alpha \hat{\bf e}(\lambda)_\beta
= \delta_{\alpha\beta} -\frac{1}{2}\{{\bf S}_\alpha,{\bf S}_\beta\} 
     +\frac{i}{2}\epsilon_{\alpha\beta\gamma}{\bf S}_\gamma \ ,
\label{eq:ee}
\end{equation}
${\bf S}$ being the spin operator. 

The final $r$-space expressions for the charge and current operators 
of the spin $1$ particle are:
\begin{eqnarray}
\rho_2({\bf q})&=&Z_2 {\rm e}^{i{\bf q}\cdot{\bf r}_2} 
        -i\,\frac{2\mu_2 -Z_2}{2 M_2^2} 
        {\bf q}\cdot({\bf S}_2\times{\bf p}_2)
             {\rm e}^{i{\bf q}\cdot{\bf r}_2} \nonumber \\
      && +\frac{Q_2}{2} {\rm e}^{i{\bf q}\cdot{\bf r}_2}
          (\frac{2}{3}|{\bf q}|^2 - ({\bf S}_2\cdot{\bf q})^2 ) \ , 
\nonumber \\ 
{\bf j}_2({\bf q})&=&
            \frac{Z_2}{2 M_2}\{{\bf p}_2,{\rm e}^{i{\bf q}\cdot{\bf r}_2}\}
            -i\,\frac{\mu_2}{M_2} ({\bf q}\times{\bf S}_2)
             {\rm e}^{i{\bf q}\cdot{\bf r}_2} \  .
\label{eq:rj2}
\end{eqnarray}
Notations are similar to the ones used in Eq.~(\ref{eq:rj1}). 
It is important to note that besides for the quadrupole moment term, 
Eq.~(\ref{eq:rj2}) and Eq.~(\ref{eq:rj1}) are formally identical.

To calculate the MM interaction between the two spin-1/2 and spin-1 particles, 
we consider the standard one-photon exchange Feynman diagram, 
from which we can write:
\begin{eqnarray}
v_{MM}(r)&=&\int d{\bf q}\,{\rm e}^{i{\bf q}\cdot{\bf r}}
v_{MM}({\bf q}) \ , \nonumber \\ 	
v_{MM}({\bf q})&=&\frac{e^2}{|{\bf q}|^2}
[\rho_1({\bf q})\rho_2({\bf q})-{\bf j}_1({\bf q})\cdot{\bf j}_2({\bf q}) ] \ .
\label{eq:vmm}
\end{eqnarray}
With a straightforward algebra, using Eqs.~(\ref{eq:rj1}) and~(\ref{eq:rj2}) 
and keeping terms up to $O(1/M^4)$, 
the formulas for $v_{MM}(Nd)$ of Eqs.~(\ref{ndMM}) and~(\ref{ndMMf}) are 
obtained.

In an equivalent derivation, $v_{MM}(Nd)$ is written as sum of the 
$NN$ MM interactions between each nucleon of the deuteron and the 
third particle, at large separation distances. 
It is however important to note that the center of mass (c.m.) of 
each two-body $NN$ subsystem is not at rest, and therefore 
Eqs.~(\ref{mmpp})--(\ref{mmnn}), which are derived in the c.m. 
reference frame, need to be generalized. In fact, 
the MM interaction between 
two spin-1/2 point-like particles 
in a generic reference frame in which the c.m. 
of the system has momentum ${\bf P}$,
is given by~\cite{stoks90,Aus83}:
\begin{eqnarray}
v_{MM}(r)&=&
-{\alpha\over r^3}\{ {\mu_1\mu_2\over M_1 M_2} S_{12}
       +\frac{Z_2}{2}\,({\mu_1\over M_1 M_{12}}-{Z_1\over 2 M^2_1})
        ({\bf L}\cdot{\bf S}+{\bf L}\cdot{\bf A})\nonumber \\
    && +\frac{Z_1}{2}\,({\mu_2\over M_2 M_{12}}-{Z_2\over 2 M^2_2})
        ({\bf L}\cdot{\bf S}-{\bf L}\cdot{\bf A}) \nonumber \\
    && -\frac{Z_1 Z_2}{4 M_1 M_2}
        [ ({\bf r}\times{\bf P})\cdot{\bf A} + 
          ({\bf r}\times{\bf P})\cdot{\bf S}\,\frac{M_2-M_1}{M_1+M_2}]\}\;\; . 
\label{eq:MM1}
\end{eqnarray}
Here $M_i$, $Z_i$, $\mu_i$ ($i=1,2$) and $M_{12}$ are the masses, charges, 
magnetic moments and reduced mass 
of the two particles, ${\bf r}$ is their 
relative position, $S_{12}=3({\bf S}_1\cdot{\hat r})
({\bf S}_2\cdot{\hat r})-{\bf S}_1\cdot{\bf S}_2$ is the tensor operator, 
${\bf S}_1$ and ${\bf S}_2$ being the spin operators, 
${\bf S}$ and ${\bf A}$ are 
defined as ${\bf S}={\bf S}_1+{\bf S}_2$ and  
${\bf A}={\bf S}_1-{\bf S}_2$, ${\bf L}$ is the orbital angular momentum.
The last term of Eq.~(\ref{eq:MM1}) is the well known Thomas precession 
(TP) term (see Ref.~\cite{For95} and references therein). 
Clearly, Eq.~(\ref{eq:MM1}) becomes Eqs.~(\ref{mmpp})-(\ref{mmnn}), 
when we consider two nucleons in their c.m. reference frame.
If the TP contribution, which is present only in $v_{MM}(pp)$ 
($Z_1\neq 0$ and $Z_2\neq 0$), was 
neglected, the $p-d$ MM interaction would have become 
\begin{eqnarray}
v_{MM}(pd)&=&-{\alpha\over r^3}[ {\mu_p\mu_d\over M_p M_d} S_{pd}^I
       +({\mu_p\over 2 M_p M_{pd}}-{1\over 4 M_p M_{pd}})
        ({\bf L}\cdot{\bf S}_{pd}+{\bf L}\cdot{\bf A}_{pd})\nonumber \\
    && +({\mu_d\over 2 M_d M_{pd}}-{1\over 4 M_d M_{pd}})
        ({\bf L}\cdot{\bf S}_{pd}-{\bf L}\cdot{\bf A}_{pd})
       -{Q_d\over 2} S^{II}_d]\;\; ,  \label{pdMMw} 
\end{eqnarray}
with same notation as in Eq.~(\ref{pdMM}).

Finally, note that Eq.~(\ref{eq:MM1}) gives the MM interactions also 
for four-body systems like $p-\,^3{\rm He}$ and $n-\,^3{\rm H}$.

\newpage

\begin{figure}
\begin{center}
\includegraphics[height=20cm]{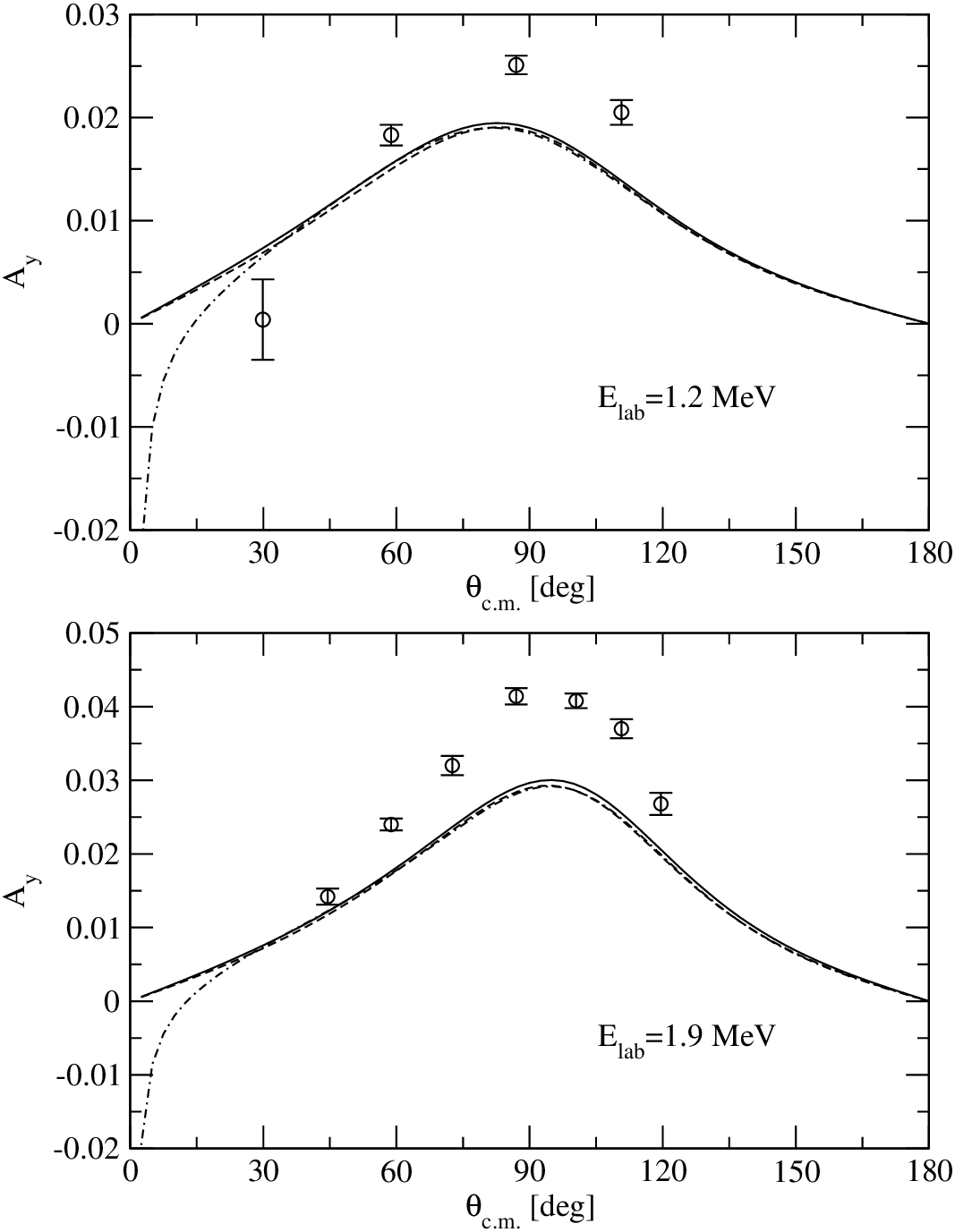}
\end{center}
\caption{The $n-d$ $A_y$ calculated using AV18 (solid line) and
AV18+MM (dotted-dashed line). For the dashed line see text. Experimental
points are from Ref.\protect\cite{werner2002}.}
\end{figure}

\begin{figure}
\begin{center}
\includegraphics[width=15cm]{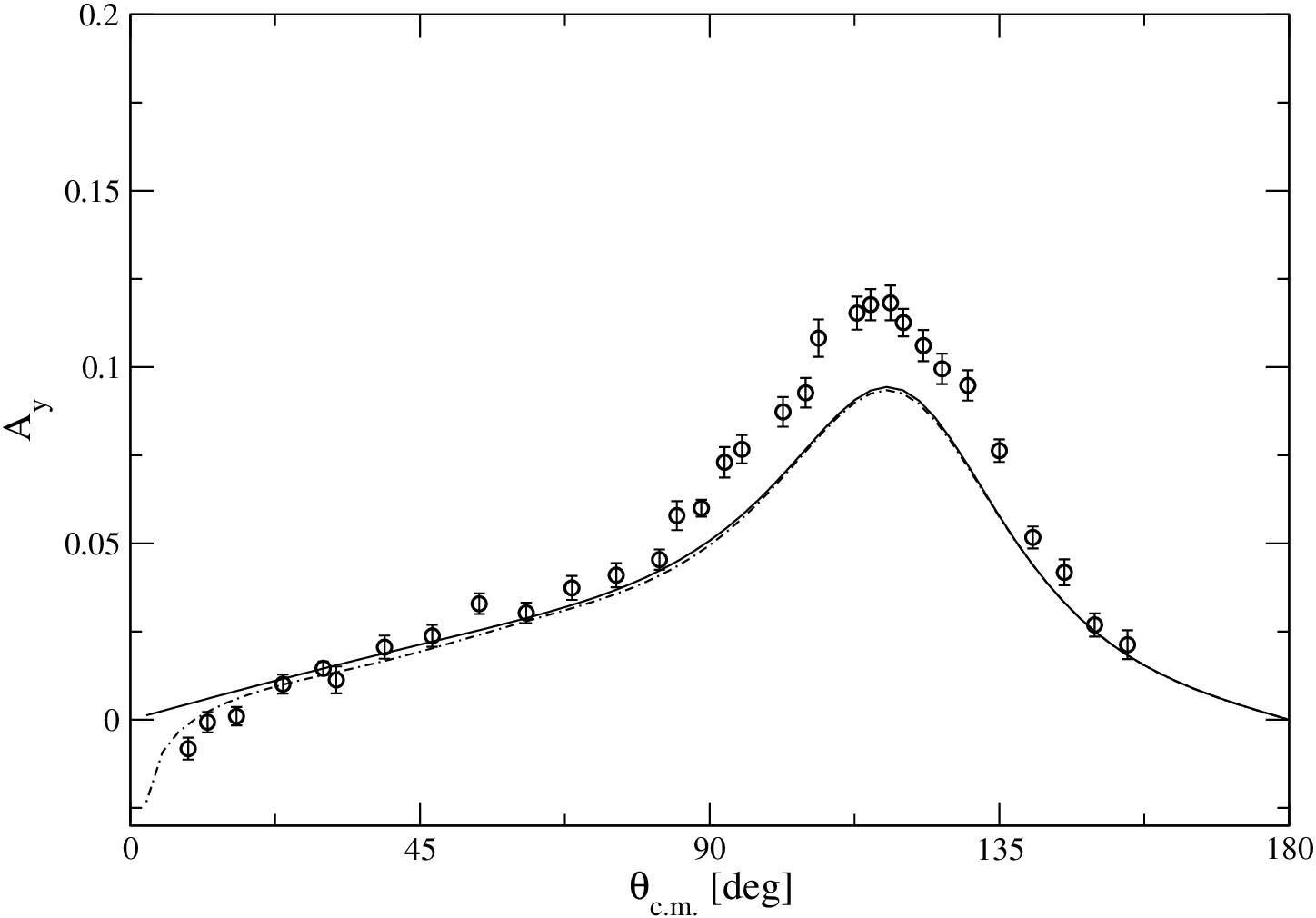}
\end{center}
\caption{The $n-d$ $A_y$ at $E_{lab}=6.5$ MeV calculated using AV18 
(solid line) and AV18+MM (dotted-dashed line). Experimental
points are from Ref.\protect\cite{werner86}.}
\end{figure}

\vspace{2cm}

\begin{figure}
\begin{center}
\includegraphics[width=15cm]{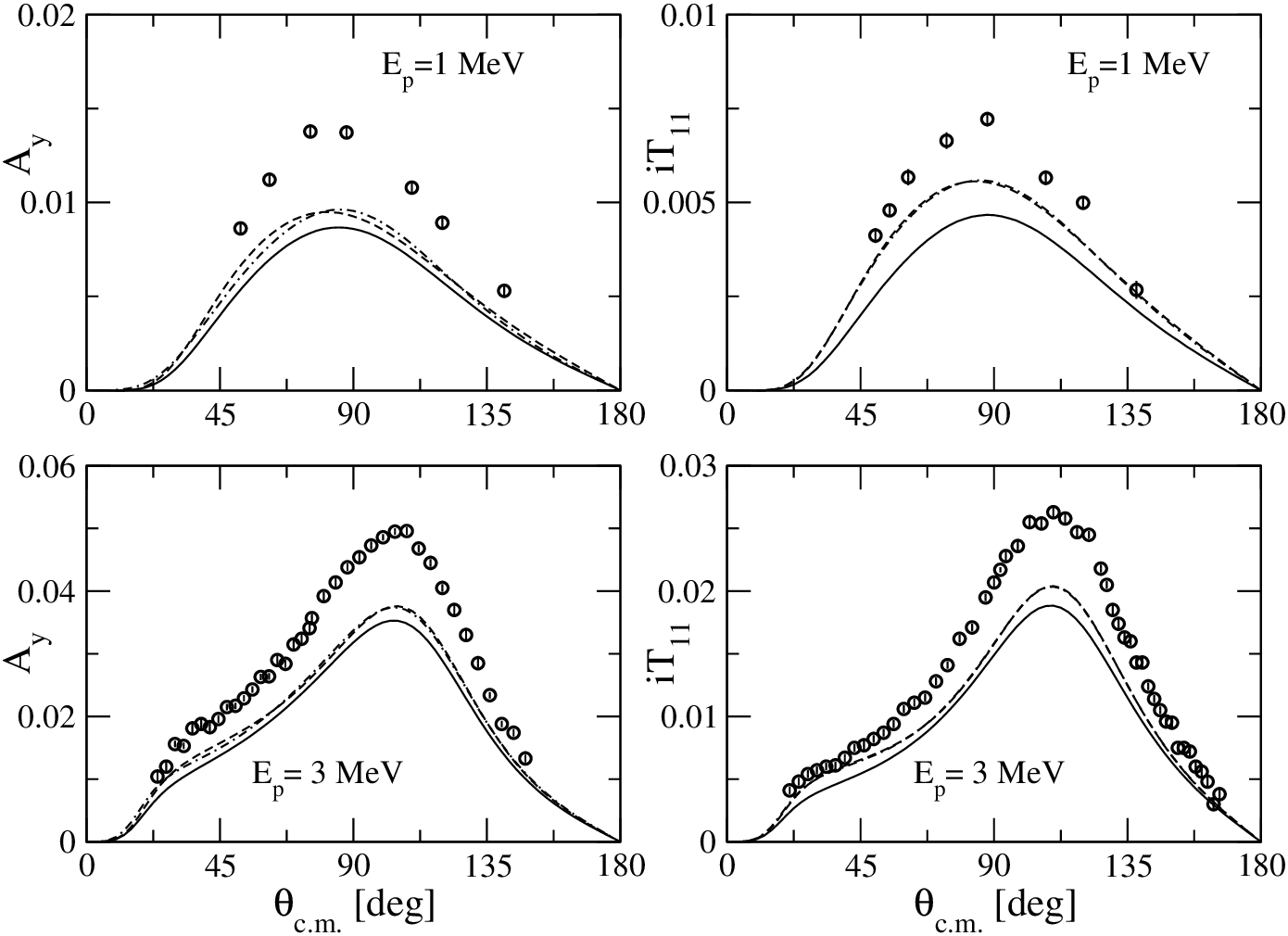}
\end{center}
\caption{The $p-d$ $A_y$ and $iT_{11}$ calculated using AV18 (solid line) and
AV18+MM (dotted-dashed line). For the dashed line see text. Experimental
points are from Ref.\protect\cite{wood} ($1$ MeV) and Ref.\protect\cite{sagara}
($3$ MeV).}
\end{figure}

\vspace{2cm}

\begin{figure}
\begin{center}
\includegraphics[width=15cm]{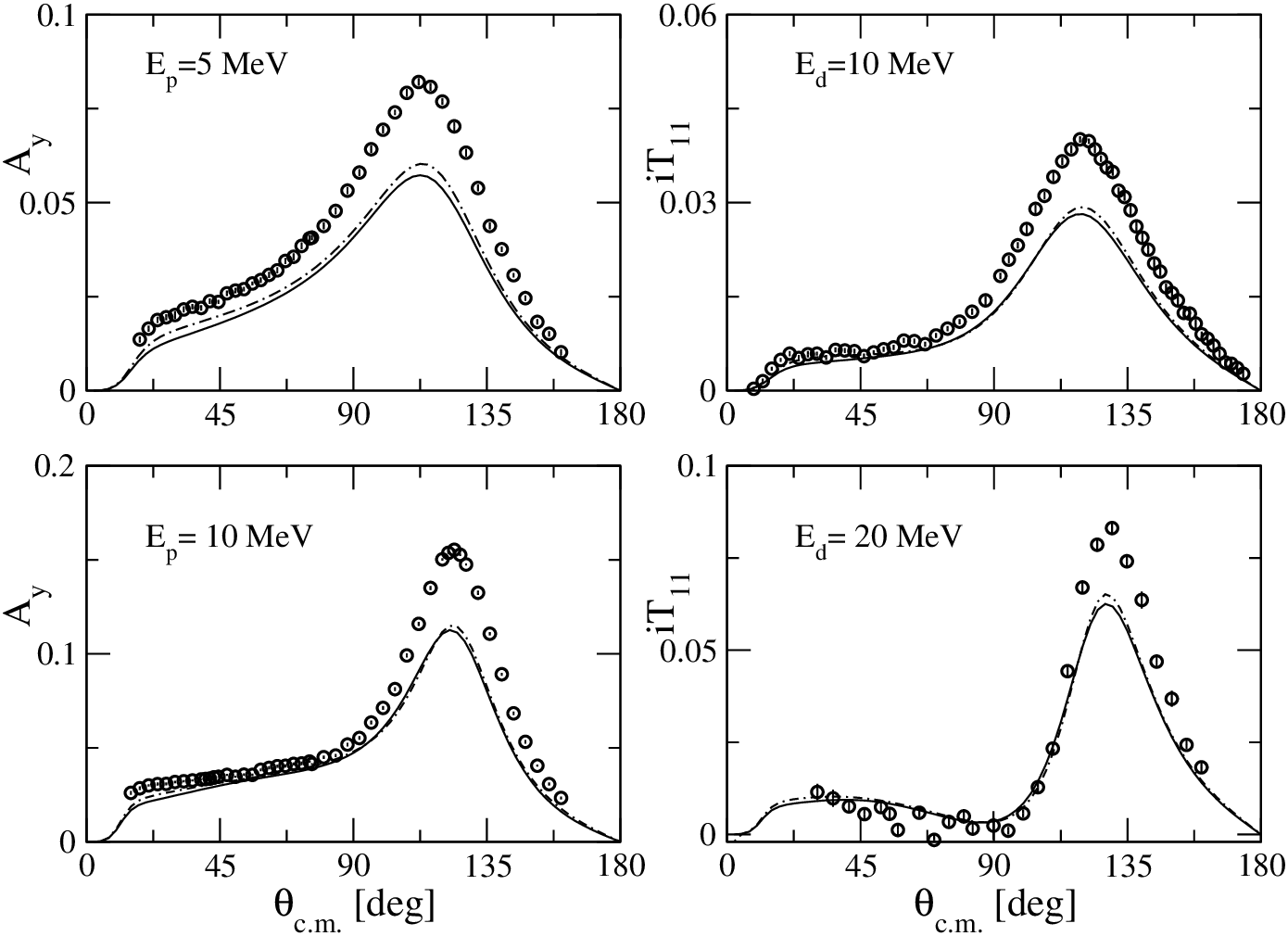}
\caption{The $p-d$ $A_y$ and $iT_{11}$ calculated using AV18 (solid line) and
AV18+MM (dotted-dashed line). Experimental
points at $E_p=5,10$ MeV and $E_d=10$ MeV are from Ref.\protect\cite{sagara1},
$E_d=20$ MeV are from Ref.\protect\cite{gruebler}.}
\end{center}
\end{figure}

\vspace{2cm}

\begin{figure}
\begin{center}
\includegraphics[width=15cm]{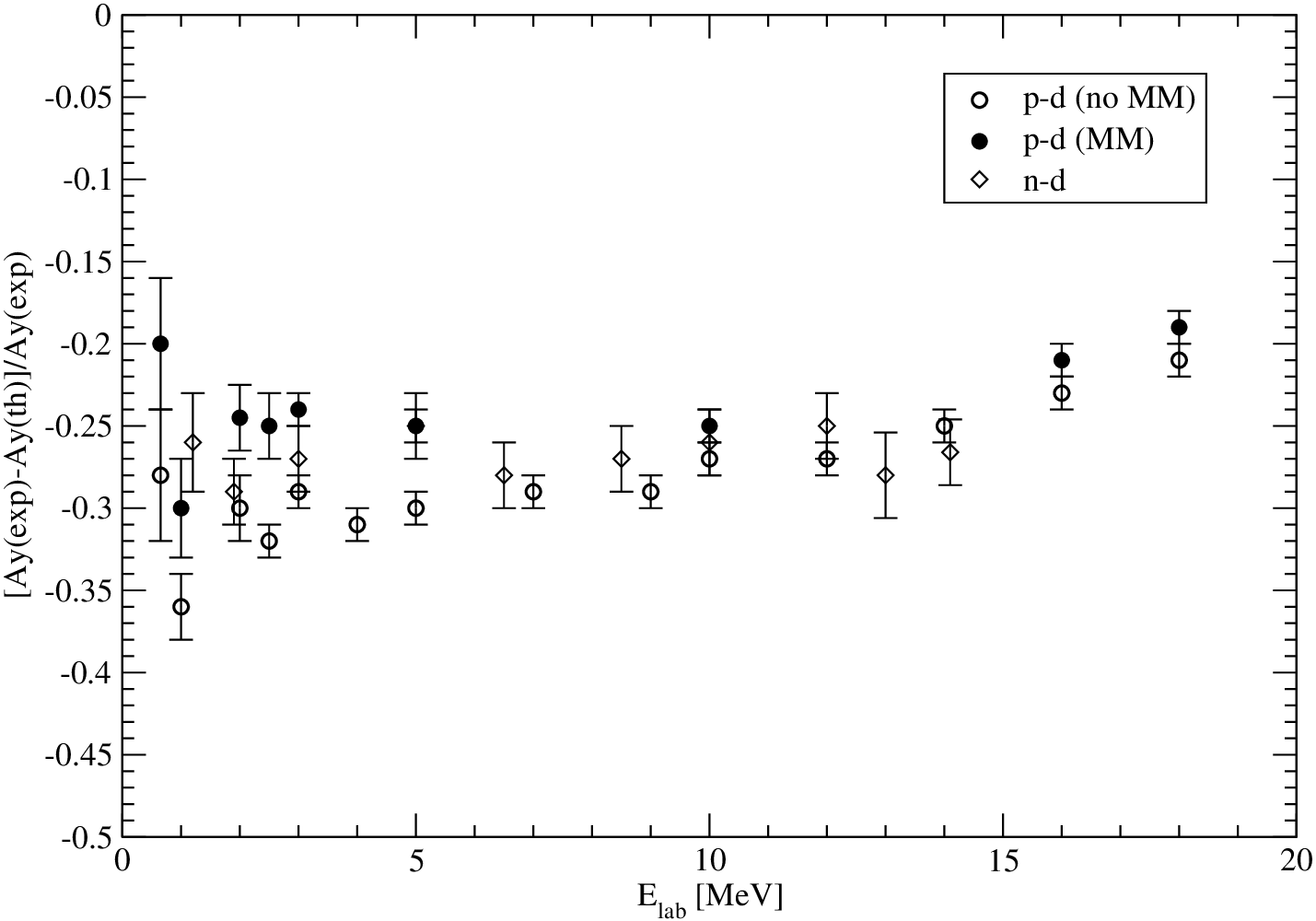}
\end{center}
\caption{Relative difference between the theoretical and experimental
vector analyzing power $A_y$ at the peak as a function of energy.}
\end{figure}

\vspace{2cm}

\begin{figure}
\begin{center}
\includegraphics[width=15cm]{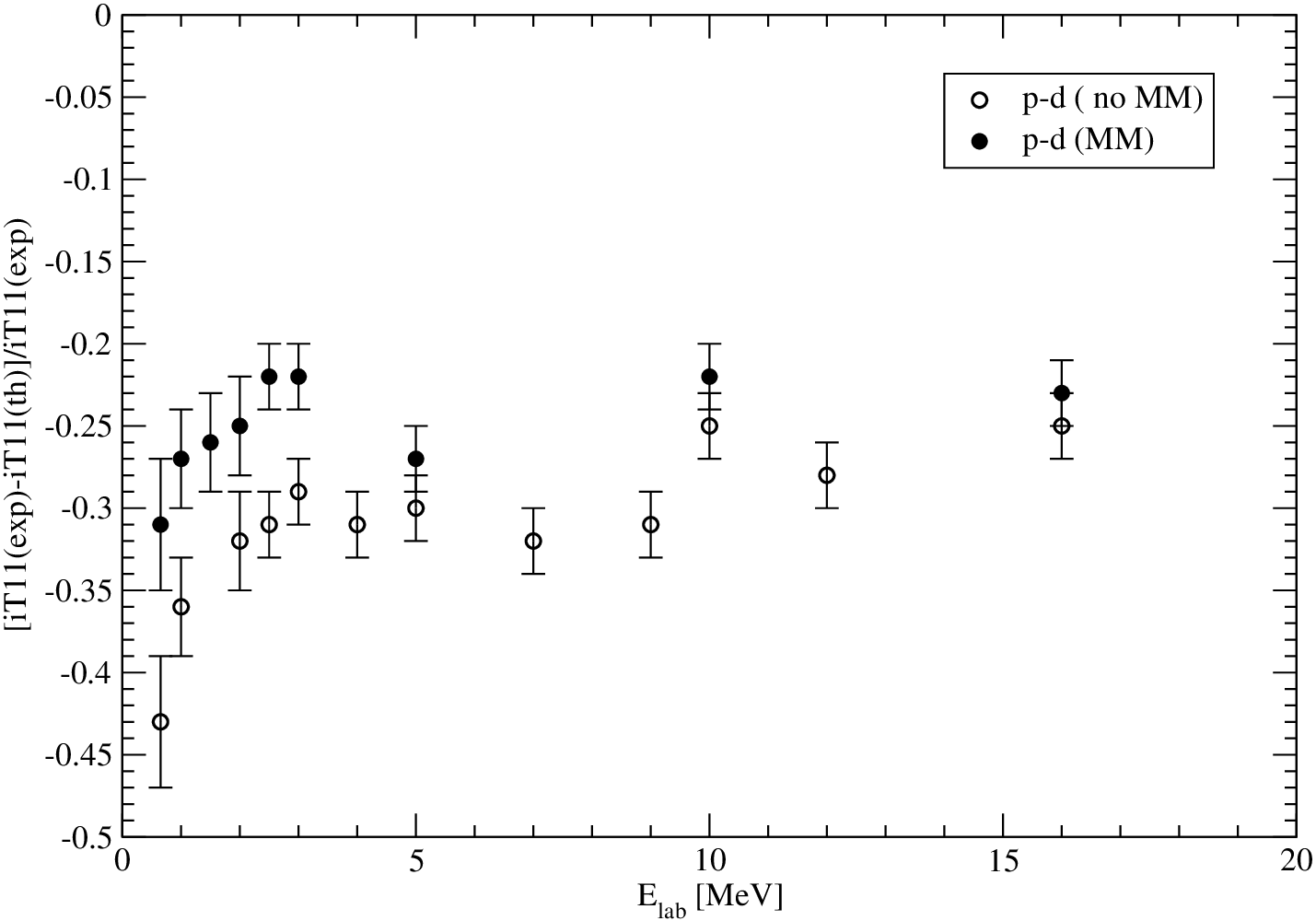}
\end{center}
\caption{Relative difference between the theoretical and experimental
vector analyzing power $iT_{11}$ at the peak as a function of energy.}
\end{figure}

\begin{figure}
\begin{center}
\includegraphics*[width=15cm]{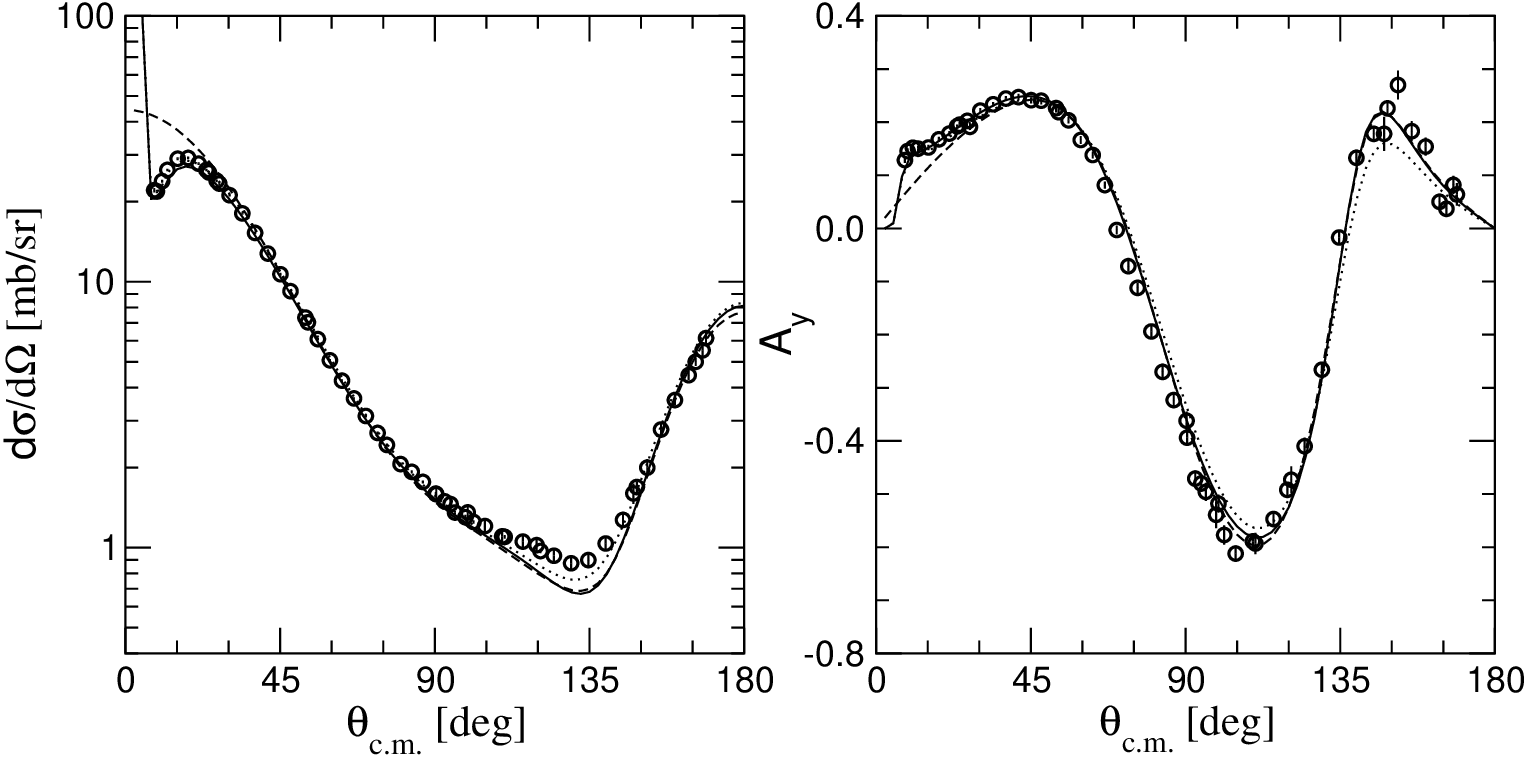}
\end{center}
\caption{ The differential cross section and $A_y$ at $E_{lab}=65$ MeV.
For explanation of the curves see text.
Experimental data are from Ref.~\protect\cite{shimizu}.}
\end{figure}

\begin{figure}
\begin{center}
\includegraphics*[width=15cm]{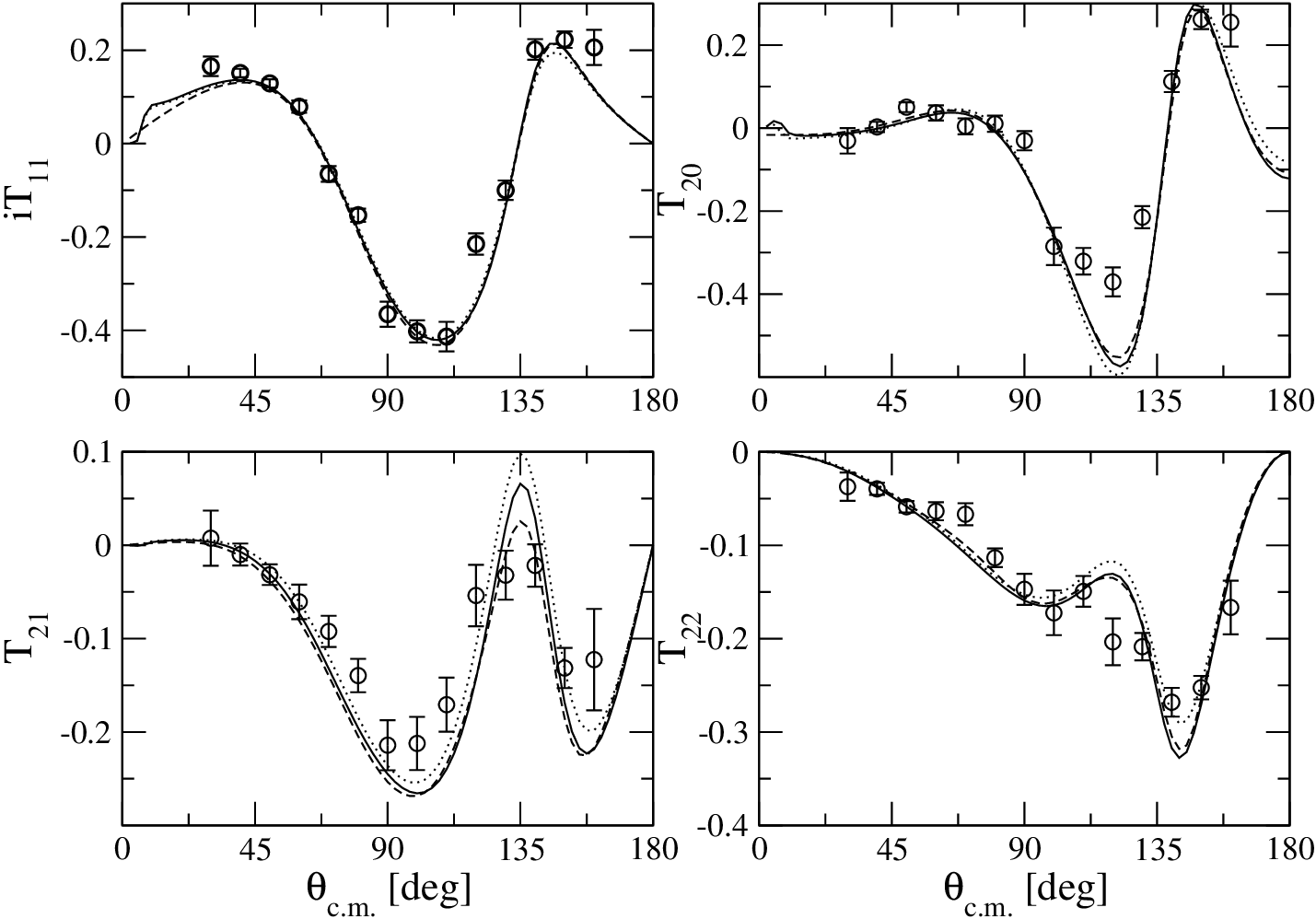}
\end{center}
\caption{The deuteron analyzing power $iT_{11}$ and the tensor analyzing
powers $T_{20},T_{21},T_{22}$ at $E_{lab}=65$ MeV.
For explanation of the curves see text.
Experimental data are from Ref.~\protect\cite{witala2}.}
\end{figure}


\begin{thebibliography}{9}

\bibitem{knutson78} L.D. Knutson and D. Chiang, Phys. Rev. {\bf{C18}}, 1958 (1978)
\bibitem{stoks90} V.G.J. Stoks and J.J. de Swart, Phys. Rev. {\bf{C42}}, 1235 (1990)
\bibitem{kvr2001} A. Kievsky, M. Viviani, and S. Rosati,
        Phys. Rev. {\bf C64}, 024002  (2001) 
\bibitem{viviani2000} M. Viviani, A. Kievsky, and S. Rosati,
        Few-Body Syst. {\bf 30}, 39 (2001)
\bibitem{phh1} A. Kievsky, M. Viviani, and S. Rosati,
              Nucl. Phys. {\bf A577 }, 511 (1994)
\bibitem{phh2} A. Kievsky, M. Viviani, and  S. Rosati,
              Nucl. Phys. {\bf A551}, 241  (1993) 
\bibitem{puzzle} A. Kievsky, S. Rosati, W. Tornow, and M. Viviani,
                 Nucl. Phys.  {\bf A607}, 402  (1996)
\bibitem{report} W. Gl\"ockle {\sl et al.}, Phys. Rep. {\bf 274}, 107 (1996)
\bibitem{witala1}H. Wita\l a {\sl et al.},
               Phys. Rev. {\bf{C63}}, 024007 (2001)
\bibitem{av18}R.B. Wiringa, V.G.J. Stoks, and R. Schiavilla,
               Phys. Rev. {\bf{C51}}, 38  (1995)
\bibitem{A8} Steven C. Pieper, V.R. Pandharipande, R.B. Wiringa, and
             J. Carlson, Phys. Rev. {\bf{C64}}, 014001 (2001)
\bibitem{nogga} A. Nogga {\sl et al.}, Phys. Rev. {\bf{C67}}, 034004 (2003)
\bibitem{stoks} V.G.J. Stoks, Phys. Rev. {\bf{C57}}, 445 (1998)
\bibitem{kie2002} A. Kievsky, M. Viviani, L.E. Marcucci, and S. Rosati,
        Few-Body Syst. Suppl. {\bf 14}, 111 (2003) 
\bibitem{hw2002} H. Wita\l a {\sl et al.},
               Phys. Rev. {\bf{C67}}, 064002 (2003)
\bibitem{bc55}L.C. Biedenharn and C.M. Class, Phys. Rev. {\bf 98}, 691
              (1955)
\bibitem{knutsonp} L.D. Knutson (private communication)
\bibitem{kohn} A. Kievsky, Nucl. Phys. {\bf{A624}}, 125 (1997)
\bibitem{seyler} R.G. Seyler, Nucl. Phys. {\bf A124}, 253 (1969)
\bibitem{werner2002}E.M. Neidel {\sl et al.}, Phys. Lett. {\bf B552}, 29 (2003)
\bibitem{werner86}W. Tornow {\sl et al.}, Phys. Lett. {\bf B257}, 273 (1991)
\bibitem{brune}C.R. Brune {\sl et al.}, Phys. Rev. {\bf C63}, 044013 (2001)
\bibitem{wood} M.H. Wood {\sl et al.}, Phys. Rev. {\bf C65}, 034002 (2002)
\bibitem{knutson} L.D. Knutson, L.O. Lamm, and J.E. McAninch,
                  Phys. Rev. Lett.{\bf 71}, 3762 (1993)
\bibitem{sagara} S. Shimizu {\sl et al.}, Phys. Rev. {\bf C52}, 1193 (1995)
\bibitem{kievwood} A. Kievsky {\sl et al.}, Phys. Rev. {\bf C63}, 024005 (2001)
\bibitem{sagara1} K. Sagara {\sl et al.}, Phys. Rev. {\bf C50}, 576 (1994);
                  K. Sagara (private communication)
\bibitem{gruebler} W. Gr\" uebler {\sl et al.}, Nucl. Phys. {\bf A398}, 445 (1983);
  F. Sperisen {\sl et al.}, {\sl ibid.} {\bf A422}, 81 (1984)
\bibitem{shimizu} H. Shimizu {\sl et al.}, Nucl. Phys. {\bf A382}, 242 (1982)
\bibitem{witala2} H. Witala {\sl et al.}, Few-Body Syst. {\bf 15}, 67 (1993)
\bibitem{urbana}B.S. Pudliner {\sl et al.}, 
                Phys. Rev. Lett. {\bf 74}, 4396 (1995)
\bibitem{Car98} J. Carlson and R. Schiavilla, 
Rev. Mod. Phys. {\bf 70}, 743 (1998)
\bibitem{Arn80} R.G. Arnold, C.E. Carlson, and F. Gross,
Phys. Rev. {\bf C21}, 1426 (1980)
\bibitem{Aus83} G.J.M. Austen and J.J. de Swart, 
Phys. Rev. Lett. {\bf 50}, 2039 (1983)
\bibitem{For95} J.L. Forest, V.R. Pandharipande, and J.L. Friar, 
Phys. Rev. {\bf C52}, 568 (1995)

\end{thebibliography}
\end{document}